\documentstyle[twoside,fleqn,espcrc2,epsf]{article}

\newcommand{\AmS}{{\protect\the\textfont2
  A\kern-.1667em\lower.5ex\hbox{M}\kern-.125emS}}

\title{SUSY and the mass difference of $B_d^0-\overline B_d^0$}

\author{G. Couture and H. K\"onig\thanks{Talk presented by H. K\"onig}
\address{D\'epartement de Physique, Universit\'e du Qu\'ebec
\`a Montr\'eal,\\ C.P. 8888, Succ. Centre-Ville, Montr\'eal,
Qu\'ebec, Canada H3C 3P8}} 
       
\begin{document}


\begin{titlepage}

\large
\begin{flushleft}
UQAM-PHE/96-07\\[0.1cm]
September 1996
\end{flushleft}
\vspace{0.4cm}
\begin{center}
\LARGE
{\bf SUSY and the mass difference of $B_d^0-\overline B_d^0$}\\
\vspace{1cm}
\large
G. Couture and H. K\"onig\\
\vspace{0.4cm}
\large {\it D\'epartement de Physique, Universit\'e du Qu\'ebec
\`a Montr\'eal,\\ 
\vspace{0.1cm}
C.P. 8888, Succ. Centre-Ville, Montr\'eal,
Qu\'ebec, Canada H3C 3P8}\\
\vspace{3cm}
{\bf Abstract}
\end{center}
\normalsize
We present an overview of the 
loop corrections to the mass difference $\Delta m_{B_d^0}/
m_{B_d^0}$\ within the minimal supersymmetric standard
model. We include the complete mixing matrices of the
charginos and neutralinos as well as of the scalar partners of
the left and right handed third generation quarks. We show
that the SUSY contribution to the mass difference in the
$B_d^0$\ system is comparable to the Standard Model one and can be even
larger for the charged Higgs contribution and for a certain supersymmetric
parameter space.

\vspace{2cm}
\noindent
\normalsize
To appear in the Proceedings of the 2nd International Conference on
Hyperons, Charm and Beauty Hadrons, Montr\'eal, Qu\'ebec, Canada,
27-30 August 1996. Edited by C.S. Kalman et al.
\end{titlepage}

\begin{abstract}
We present an overview of the 
loop corrections to the mass difference $\Delta m_{B_d^0}/
m_{B_d^0}$\ within the minimal supersymmetric standard
model. We include the complete mixing matrices of the
charginos and neutralinos as well as of the scalar partners of
the left and right handed third generation quarks. We show
that the SUSY contribution to the mass difference in the
$B_d^0$\ system is comparable to that of the Standard Model and can be even
larger than that of the charged Higgs for parts of the supersymmetric
parameter space.
\end{abstract}

\maketitle

\section{Introduction}

The mass difference $\Delta m_{B_d^0}/m_{B_d^0}\approx
5.9\times 10^{-14}$\ GeV ~\cite{prd} is an experimentally well known
value. In the standard model (SM), where
the W bosons and up-type quarks run in the
relevant box diagrams, it was found that
in the $B_d^0$\ system the top quark leads to the most
important contribution.

Nowadays, from CDF and D$\emptyset$ we know that the top quark mass 
is about 180 GeV~\cite{cdfdo}. 
Because of such a large value, we have to reconsider
the influence of one of the most favoured models beyond the
SM, its minimal supersymmetric extension (MSSM)~\cite{nil,habkan}, 
to $\Delta m_{B_d^0}$. As is well known, the particle spectrum of the
MSSM is enhanced by at least a factor of two and therefore many
more particles contribute to the mass difference of the 
$B_d^0$\ system. 

In this talk we present an overview of the contribution of all
particles within the MSSM to this electroweak parameter. We
present the results only in a very general way with a short
discussion and refer the interested reader to our articles
~\cite{gchk1,gchk2} for more detailed and complete calculations. 

\section{The SM and $\Delta m_{B_d^0}$}

In the SM, the contribution of the W bosons
and up-type quarks to the mass difference of
the $B_d^0$\ system is given by ~\cite{busloste}:\footnote{ 
The explicit forms of 
eqs.(\ref{smeq}-\ref{higeq})  are quite 
lenghty and can be found in~\cite{gchk1,gchk2}}

\begin{equation}
\label{smeq}
{{\Delta m_{B_d^0}}\over{m_{B_d^0}}}=
{{G_F^2}\over{6\pi^2}}f^2_BB_B\eta_tm^2_W
(K_{31}^\ast K_{33})^2S(x_t) 
\end{equation}

$f_B$, $B_B$\ are the structure constant and the
Bag factor obtained by QCD sum rules and $\eta_t$\
a QCD correction factor.

\section{The MSSM and $\Delta m_{B_d^0}$}

The first obvious thing to do to obtain $\Delta m_{B_d^0}$\
within the MSSM is to replace the W bosons by their fermionic
partners and the up quarks by their scalar partners. Unfortunately
due to mass mixing effects and 
the rich particle spectrum of the MSSM things are not quite
that simple.

First of all due to new scalar fermion-fermion-gaugino couplings
the fermionic partners of the W bosons mix with the fermionic
partners of the charged Higgses when the neutral scalar Higgses
obtain their vacuum expectation values (vev). The mass eigentstates
are known as charginos.

Second, the scalar partners of the left and right handed
up-type quarks will mix. This mixing is proportional to the up-type 
quark masses; since  
the top quark mass is very large it 
cannot be neglected for the third generation. Although the
mixing of the scalar partners of the left and right handed bottom
quarks is proportional to the bottom quark mass, we did not
neglect it since for large values of $\tan\beta=v_2/v_1$\ 
(the ratio of the vevs) it can become important too.

As a result, after a lengthy but straightforward calculation we obtain
the following result when charginos and scalar up quarks
are running in the loop:

\begin{eqnarray}
\label{cheq}
{{\Delta m_{B_d^0}}\over{m_{B_d^0}}}&=&
{{G_F^2}\over{4\pi^2}}f^2_BB_Bm^4_W(K^\ast_{31}K_{33})^2
\\ \nonumber
& &\lbrack Z^{\tilde W}_{11}-2 Z'^{\tilde W}_{31}+
\tilde Z^{\tilde W}_{33}\rbrack
\end{eqnarray}

As we can see from the structure of eq.(\ref{cheq}) if 
all the scalar quark masses are degenerate, the result
is identically 0 (GIM mechanism). Since we neglected all
quark masses beside the top and bottom quark masses, we made use
of $Z_{11}=Z_{12}=Z_{21}=Z_{22}$\ and $Z_{13}=Z_{31}=Z_{32}=
Z_{23}$. In this case, only the mass difference between the first and
third generations of the scalar quarks comes into play.

It was shown more than 10 years ago that loop diagrams induce
flavour changing couplings of the gluinos (the fermionic partners
of the gluons) to the down quarks and their scalar partners
~\cite{dun,don}. Since the gluinos couple strongly their
contribution to $\Delta m_{B^0_d}$\ was thought to be the
most important one. The result is given by:

\begin{eqnarray}
\label{gleq}
{{\Delta m_{B_d^0}}\over{m_{B_d^0}}}&=&
-{{\alpha_s^2}\over{54}}f_B^2B_B(K^\ast_{31}K_{33})^2
\nonumber \\
& &\lbrack Z^{\tilde g}_{11}-2 Z'^{\tilde g}_{31}+
\tilde Z^{\tilde g}_{33}\rbrack
\end{eqnarray}

In the MSSM there are also the neutralinos (the mass
eigenstates of the fermionic partners of the photon, the
Z boson and the neutral Higgses) and the scalar partners
of the down quarks within the relevant box diagrams. Since
the neutralinos couple ony weakly their contribution has been  
neglected in the literature
so far. We show that this is illegitimate for a certain
range of supersymmetric parameter space. The calculation is very
lengthty and we obtain:

\begin{eqnarray}
\label{neeq}
{{\Delta m_{B_d^0}}\over{m_{B_d^0}}}&=&
{{G_F^2}\over{(4\pi)^2}}f^2_BB_Bm^4_Z(K^\ast_{31}K_{33})^2
\nonumber \\
& &\lbrack Z^{\tilde N}_{11}-2 Z'^{\tilde N}_{31}+
\tilde Z^{\tilde N}_{33}\rbrack
\end{eqnarray}

Finally we also want to comment on
the charged Higgs boson
contribution to the mass
difference in the $B_d^0$\ system. In the case where we neglect 
the bottom quark mass, that is
$m_b\tan\beta\ll m_t\cot\beta$\ we obtain:

\begin{eqnarray}
\label{higeq}
\frac{\Delta m_{B_d^0}}
{m_{B_d^0}}&=&{{G^2_F}\over{16\pi^2}}m^4_t
\cot^4\beta f^2_BB_B(K_{31}^\ast K_{33})^2
\nonumber \\
& &\bigl\lbrace \tilde F^{tt}_{H^+H^+} 
+2\tan^2\beta\lbrack\tilde F^{tt}_{H^+W^+}
\\ \nonumber
& &+4(m_W/m_t)^2 
_M\tilde F^{tt}_{H^+W^+}\rbrack\bigr\rbrace
\end{eqnarray}

When one has $m_b\tan\beta\sim m_t\cot\beta$\
one should not neglect the bottom quark
mass when calculating the box diagram; this complicates
greatly the calculations.

\section{Discussions}

We now present those contributions  for different values of gaugino,
gluino and scalar quark masses and charged Higgs masses
as well as the bilinear Higgs mass term
$\mu$.
We also vary $\tan\beta$ and the symmetry-breaking scales
$m_S$.

\begin{figure}[hbtp]
\begin{center}
\mbox{\epsfysize=46mm\epsffile{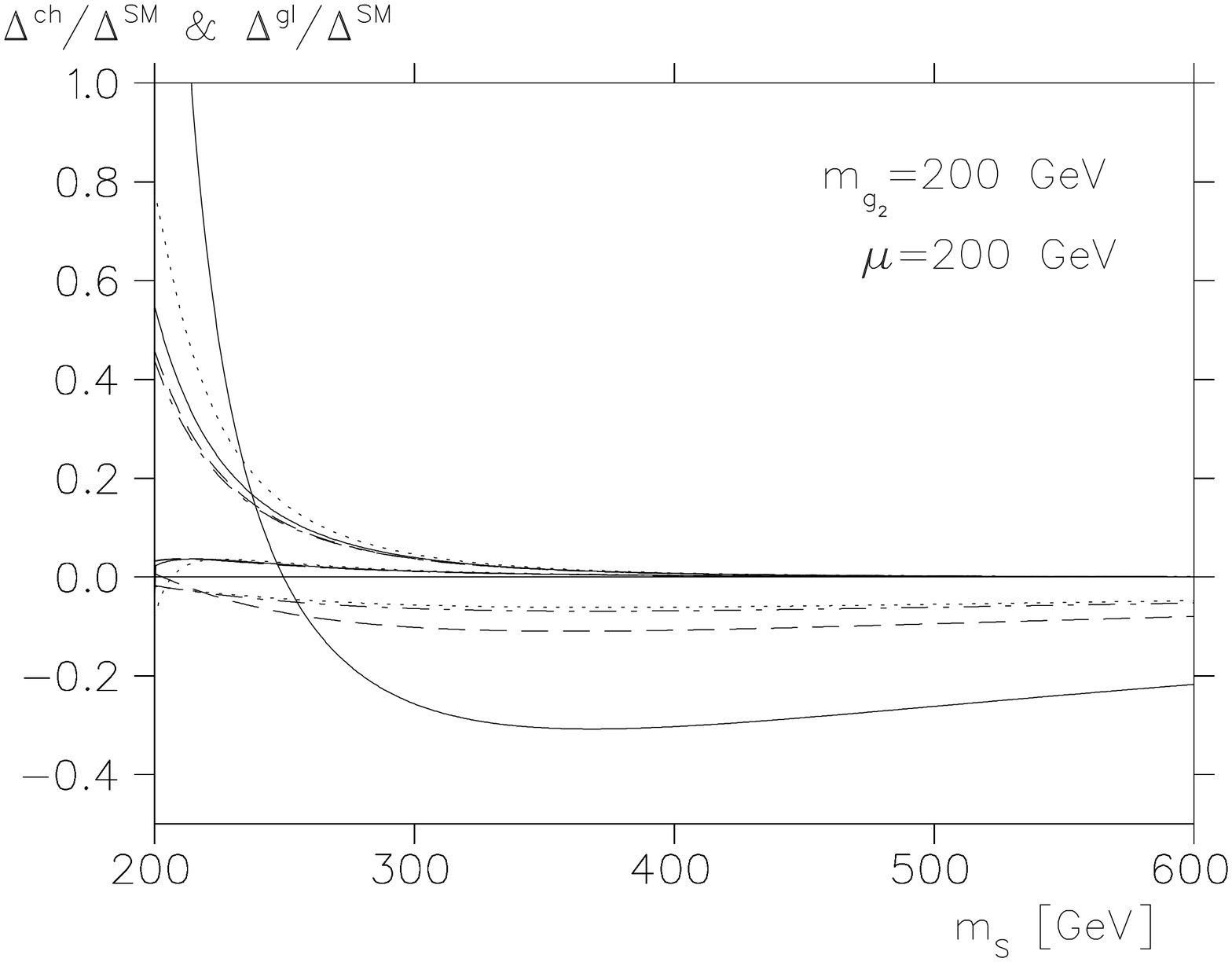}}
\end{center}
\caption{
The ratios
$\Delta m_{B^0_d}^{\rm Chargino}/\Delta m_{B^0_d}^{\rm SM}$\ and
$\Delta m_{B^0_d}^{\rm Gluino}/\Delta m_{B^0_d}^{\rm SM}$\
as a function of the scalar mass $m_S$ for $\tan\beta = 1$ (solid);
$\tan\beta = 2$ (dash); $\tan\beta = 5$ (dash-dot); $\tan\beta = 20$ (dot).
The negative values for large $m_S$ are the chargino contributions; those of
large amplitudes for small $m_S$ are the gluino contributions with
$m_{\tilde g} = 100~GeV$;
those of small amplitudes for small $m_S$ are the gluino contributions with
$m_{\tilde g} = 200~GeV$.}
\label{chglfig}
\end{figure}

In Fig.~\ref{chglfig}, we show the chargino and gluino contributions. 
The global 
behaviour is clear: for small gluino mass and small values of $m_S$, the
gluino contribution is important no matter what values the other parameters
have. On the other hand, for large gluino mass and/or large values of $m_S$,
the chargino contribution vastly dominates. The only exception to this rule is
for very large values of $\tan\beta$ ($\sim$ 30 or higher). 
This is due to the fact that such large values of $\tan\beta$\ can
push down the mass of one of the scalar b-quark eigenstates; well
below the scalar top-quark eigenstates.

The effects of the 
mixing of the scalar partners with the top and bottom quarks are more 
important for
large values of $m_S$: the contributions from the charginos don't decrease as
quickly with the mixing. For small values of $m_S$, there is also an 
enhancement. 

\begin{figure}[hbtp]
\begin{center}
\mbox{\epsfysize=45mm\epsffile{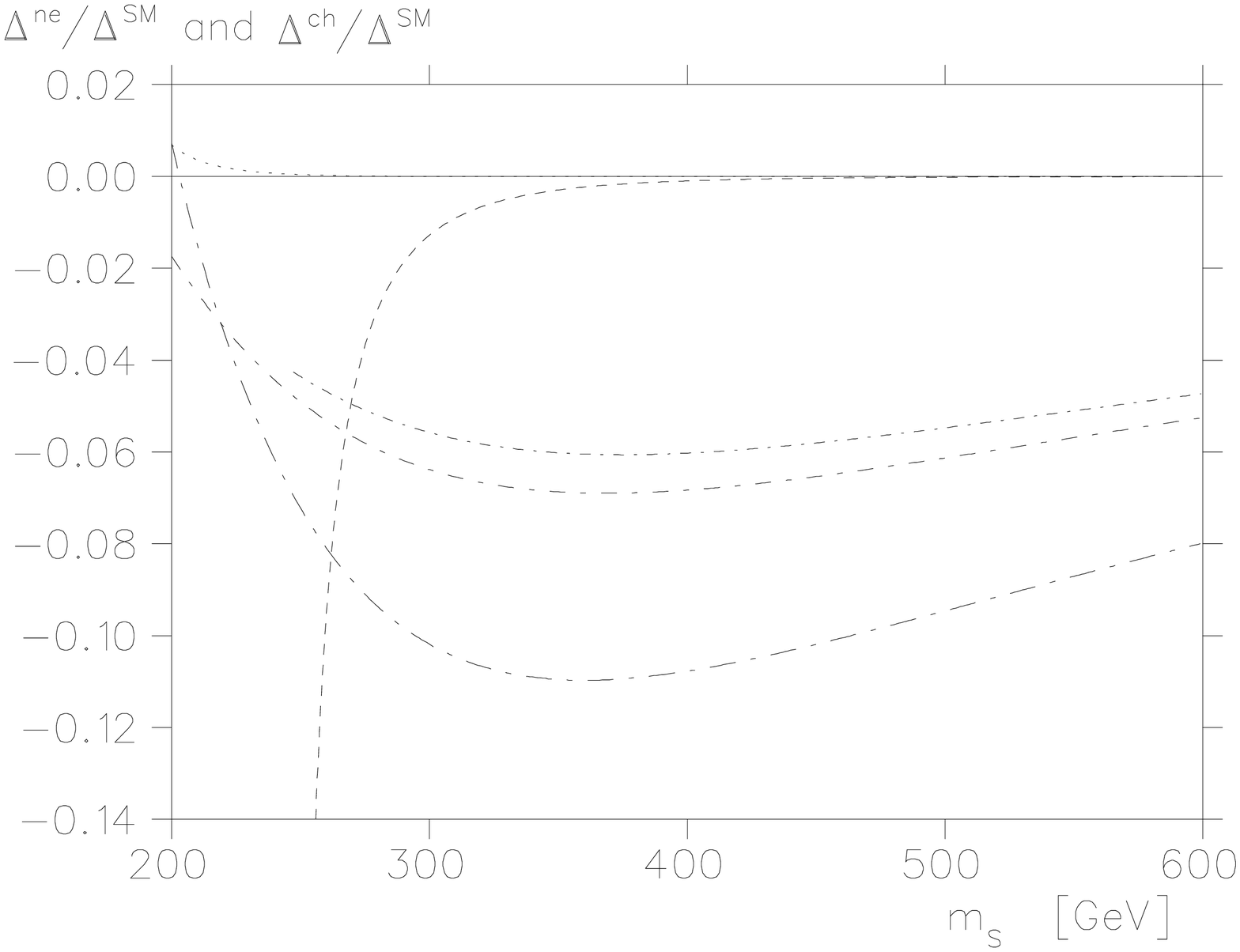}}
\end{center}
\caption{
The ratios
$\Delta m_{B^0_d}^{\rm Neutralino}/\Delta m_{B^0_d}^{\rm SM}$\ for
$\tan\beta = 2,~5$ (dotted line, the two lines are on top of each other) and
$\tan\beta = 50$ (dash)
and $\Delta m_{B^0_d}^{\rm Chargino}/\Delta m_{B^0_d}^{\rm SM}$\
for
$\tan\beta = 2$ (very long dash-dot), $\tan\beta = 5$ (long dash-dot), and
$\tan\beta = 50$ (dash-dot) as a function of $m_S$ with
$m_{g_2} = \mu = 200~GeV$.}
\label{nefig}
\end{figure}

In Fig.~\ref{nefig}, we show the neutralino contribution and 
compare them with the chargino contribution. 
The global
behaviour is clear: for small values of $\tan\beta$ ($\sim 20$ or less) the
neutralino contribution is small compared to that of the chargino. On the 
other hand, when $\tan\beta\sim 50$
\footnote{Such high values for $\tan\beta$\ are {\bf preferred}
in models, which require the Yukawa couplings $h_t,\ h_b\ 
{\rm and}\ h_\tau$\ to meet at one point at the unification scale},  the 
neutralino contribution can be much
larger than that of the chargino for the smallest possible values of 
$m_S$.  Unfortunately, as we can see on the figure, this contribution falls 
very quickly as $m_S$ increases and becomes negligible as soon as $m_S$ is a 
few tens of GeV's above its minimal value 
(for smaller values of $m_S$\ the square of one of the mass eigenvalues of the
scalar bottom quark becomes negative).

\begin{figure}[hbtp]
\begin{center}
\mbox{\epsfysize=46mm\epsffile{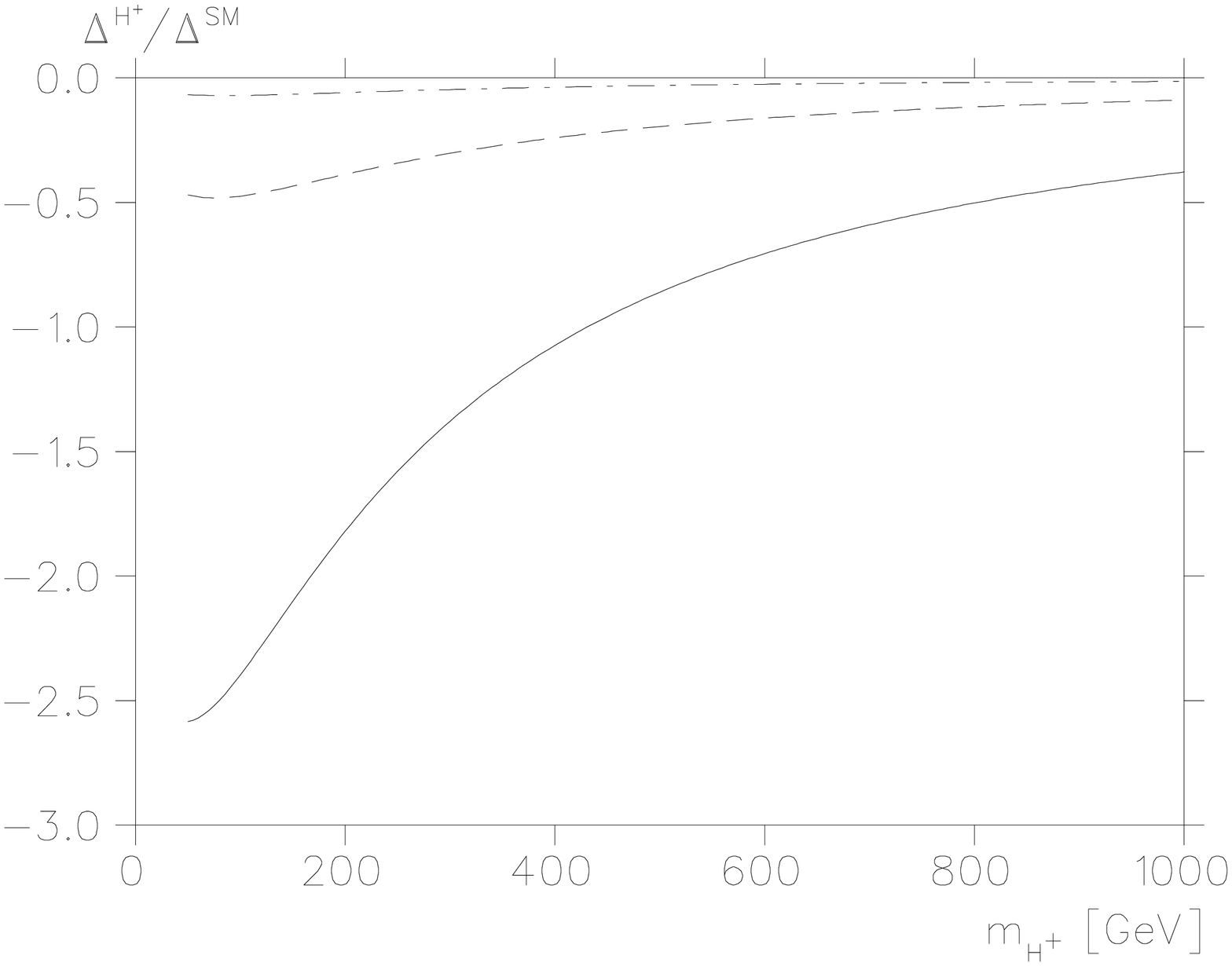}}
\end{center}
\caption{The ratio of the total amplitude
$\Delta m_{B^0_d}^{\rm H^+}/\Delta m_{B^0_d}^{\rm SM}$\ as a function
 of $M_{H^+}$ for $\tan\beta = 1$ (solid);
$\tan\beta = 2$ (dash); $\tan\beta = 5$ (dash-dot).}
\label{higgsfig}
\end{figure}

Finally in Fig.~\ref{higgsfig} we show 
the charged Higgs contribution. We see that for small
values of $\tan\beta$ and light Higgs, this contribution can exceed that of
the SM. For $\tan\beta = 1$, even for very large Higgs masses, this 
contribution is still 20\% of the SM contribution. However, this contribution
goes down very quickly when $\tan\beta$ increases. Given our 
approximation ($m_b\tan\beta\ll m_t\cot\beta$) we cannot exceed
$\tan\beta\sim 5$\ and still trust our result.

Last but not least one must not forget that in 
eqs.(\ref{cheq}-\ref{neeq})
$K_{31}^\ast K_{33}$\ have not necessarily the same
values as in the SM.
The Kobayashi--Maskawa matrix in the couplings of the
charginos to quarks and scalar quarks is multiplied by
another matrix $V_u$, which can be parametrized
by $\epsilon_u$,  
so that $K\equiv V_u\cdot K_{SM}$. 
If $\epsilon_u\ll 1$ then $K\sim K_{SM}$.
However, with
$\epsilon_u=0.3$, $K^\ast_{31}K_{33}$\ is enhanced by a
factor of 3 over the SM value. 

We have a similar matrix in the gluino--down quark--scalar
down quark couplings parametrized by $\epsilon_d$.
For $\epsilon_d=0.1$, $K^\ast_{31}K_{33}$ is
identical to the SM values, whereas $\epsilon_d=0.3$\ 
enhances it by 9.

Considering that these values are to be squared in the
mass difference of the $B^0_D$\ system we can use that
enhancement to put limits on $\varepsilon_{u,d}$.
In the case at hand, $\varepsilon_u$\ has to be smaller than $0.2$\ and
$\varepsilon_d$\ smaller than $0.1$\ to keep the results
lower than the measured value of $\Delta m_{B^0_d}/m_{B^0_d}$. 
This is not very constraining yet but it is already better than the limit one 
can get from current data on rare Kaon decays \cite{gchk3} 

\section{Conclusion}

In this talk we presented the  
contributions of all particles within the MSSM  
to the mass difference in the $B_d^0$\ 
system via box diagrams. In the calculatiuons we 
included the mixing of the charginos and neutralinos and
the mixing of the scalar top and bottom quarks.

We have shown that for reasonable values of the SUSY parameters
the contribution of the box diagrams with charginos and scalar
up quarks can be of the same order as 
those of the SM diagrams, but with opposite sign. 
The same goes for the contribution of the gluino and
scalar down quarks box diagrams, which has  the same sign as
the SM contribution. 

We have shown that in the case of charginos and
scalar top quark, the mixing becomes important and leads to
an enhancement. In the case of
gluinos and scalar bottom quark the mixing is
less important and even for higher values of $\tan\beta$\ the results
are reduced only by a few per cent.

Since we have shown that despite the smallness of 
the weak coupling constant compared to the strong coupling
constant charginos and scalar up quarks cannot be
neglected, we included the contribution of the 
neutralinos and scalar down quarks and showed
that it is in general small but can be important for large
values of $\tan\beta$\ ($\sim 50$) and the smallest
possible values of $m_S$, given $m_{g_2}$\ and $\mu$.

The contribution from the charged Higgs boson
to the mass
difference in the $B_d^0$\ system can be very important for small values 
of $\tan\beta$ and small Higgs masses. When the Higgs mass becomes large 
($\sim$ 500 GeV) and/or $2\le \tan\beta$, this contribution becomes small and
even negligible compared to the chargino contribution.

We hope that our study will guide experiments at  
the upcoming {\it B-factories}.

\section{Acknowledgements}

This research was funded by NSERC of
Canada and FCAR du Qu\'ebec. H.K. would like to thank
M. Boyce for his help in all kinds of computer/Latex questions.

\end{document}